
\input phyzzx
\overfullrule = 0pt

\def\ie{{\it i.e.}}

\def\text{\textstyle}

\def\refmark#1{[#1]}                        

\magnification=\magstep1
\vsize=23.5 true cm
\hsize=15.4 true cm
\predisplaypenalty=0
\abovedisplayskip=3mm plus6pt minus 4pt
\belowdisplayskip=3mm plus6pt minus 4pt
\abovedisplayshortskip=0mm plus6pt
\belowdisplayshortskip=2mm plus6pt minus 4pt
\normalbaselineskip=12pt
\normalbaselines
\noindent

\REF\lan{ Landau, L.D., {\sl Fundamental Problems}, in { \it Pauli Memorial
Volume}, pg.  245,
(Interscience, New York, 1960). }

\REF\fey{Feynman, R., in {\it The Quantum Theory of Fields -- The $12^{\rm th}$
Solvay Conference},
1961 (Interscience, New York).}

\REF\lanpom{ Landau, L.D. and I.Pomeranchuk, {\it Dokl. Akad. Nauk SSSR} {\bf
102}, 489 (1955). }

\REF\lanft{ Landau, L.D., in { \it Niels Bohr and the Development of Physics},
 (ed. W.Pauli), p. 52. (Pergamon Press, London, 1955).}
,
\REF\John{ Johnson, K., {\sl The Physics of Asymptotic Freedom }, in
{\it  Asymptotic Realms of Physics }, ed. A. Guth (MIT Press, 1983). }

\REF\Van{Terent'ev, M.V. and Vanyashin, V.S., {\sl   The Vacuum
Polarization of a Charged Vector Field}, {\it Soviet Physics JETP \bf 21}, 380
(1965).}

\REF\smat{Chew, G.  in {\it S-Matrix Theory}, (W.A. Benjamin Inc, 1963).}

\REF\chew{Chew, G.  in {\it The Quantum Theory of Fields:
The 12th Solvay Conference}, 1961 (Interscience, New York).}

\REF\Gold{Goldberger, M.  in {\it The Quantum Theory of Fields: The 12th Solvay
Conference}, 1961 (Interscience, New York).}

\REF\low{Low, F., in {\it    Proceedings of the International Conference on
High
 Energy Physics 1966},   pg. 249 (1966).}

\REF\loww{Low, F., {\it Suppl. al Nuovo Cimento \bf Vol. IV,2},379 (1964).  }

\REF\nal{Gell-Mann, M., {\sl   Current Algebra: Quarks and  What Else?}, in
{\it
 Proceedings of the XVI Int'l Conference on High Energy Physics}, {\bf Vol. 4},
135 (1972). }

\REF\Eight{Gell-Mann, M. and Neeman, Y. {\sl The Eightfold Way}
(W.A. Benjamin Inc.) (1964).}

\REF\veal{Gell-Mann, M., {\sl    The Symmetry Group of Vector
and Axial Vector Currents}, {\it Physics, \bf1}, 63  (1964).}

\REF\quarks{ Gell-Mann, M., {\sl A Schematic Model of Baryons and Mesons},
{\it Phys. Lett. \bf 8}, 214 (1964);
Zweig, G., {\it CERN Report No. TH401}, 4R12 (1964) (unpublished).}

\REF\green{Greenberg, O.W., {\sl   Spin and Unitary-Spin Independence in a
Paraquark Model of  Baryons and Mesons}, {\it Phys. Rev. Lett. \bf 13},  598
(1964).}

\REF\hannambu{Han, M.Y. and Nambu,Y., {\sl   Three-Triplet Model with Double
SU(3) Symmetry}, {\it Phys. Rev.
\bf 139B}, 1006 (1965).}

\REF\Dashen{Dashen, R. and Gell-Mann, M., {\sl Representations of
Local Current Algebra at  Infinite Momentum}, {\it Phys. Rev. Lett.  \bf 17},
340
(1966).}

\REF\adwe{Adler, S., {\sl Sum Rules for the
Axial Vector Coupling Constant Renormalization in
$\beta $ Decay}, {\it Phys. Rev. \bf 140 B}, 736 (1965);
Weisberger,W.I. {\sl Unsubtracted Dispersion Relations and the
Renormalization of the Weak Axial-Vector Coupling Constants},
{\it Phys. Rev. \bf 143 B}, 1302 (1966).}

\REF\bbj{Bjorken, J.D.  {\sl Applications of the Chiral U(6) $\times$ U(6)
Algebra of Current Densities}, {\it Phys. Rev. \bf 148 B}, 1467 (1966).}

\REF\bj{Bjorken, J.,  {\sl Current Algebra at Small Distances},
{\it Varenna School
Lectures, Course XLI, Varenna, Italy} (1967).}

\REF\bjj{Bjorken, J.,  {\sl Inequality for Backward Electron and Muon-Nucleon
Scattering at High Momentum Transfer},  {\it Phys. Rev. \bf 163}, 1767 (1967).}

\REF\sug{Sugawara, H., {\sl A Field Theory of Currents},  {\it Phys. Rev. \bf
170 }, 1659 (1968).}

\REF\grocalsug{Callan, C.G. and Gross, D. J.,
{\sl Crucial Test of a Theory of Currents} {\it Phys. Rev.
Lett. \bf 21}, 311 (1968).}

\REF\bjjj{Bjorken, J.,  {\sl Asymptotic Sum Rules at Infinite Momentum},  {\it
Phys. Rev. \bf 179}, 1547  (1969). }

\REF\bloom{Bloom, E.D. et al., {\sl   High-Energy Inelastic $e-p$ Scattering at
$6^o$ and $10^o$}, {\it Phys. Rev. Lett. \bf 23},  930  (1969).}

\REF\grcal{Callan, C. G. and Gross, D. J., {\sl High-Energy Electroproduction
and
 the Constitution of the Electric Current}, {\it Phys. Rev. Lett. \bf 22}, 156
(1968).}

 \REF\grlwsm{Gross, D.J. and Llewelyn-Smith,C.,
{\sl High Energy Neutrino-Nucleon Scattering, Current Algebra and Partons},
{\it Nucl. Phys. \bf B14}, 337 (1969).}

\REF\parton{ Feynman, R. P., {\sl Very High-Energy Collisions of Hadrons}, {\it
Phys. Rev. Lett. \bf 23}, 1415 (1969).}

\REF\drell{ Drell, S.D. and Yan, T.M., {\sl Partons and Their Applications at
High Energies}, {\it Ann. Phys.(NY) \bf 66}, 578 (1971).}

\REF\grws{ Gross, D.J. and Wess, J., {\sl Scale Invariance,
Conformal Invariance and the High Energy Behavior of
Scattering Amplitudes}, {\it Phys. Rev. \bf D2}, 753 (1970).}

\REF\Ven{Veneziano, G., {\sl Construction of a Crossing-Symmetric Regge-Behaved
Amplitude for Linearly Rising Trajectories},  {\it Nuovo Cimento, {\bf 57A}},
190 (1968).}

\REF\thooft{ 't Hooft, G., {\sl Renormalizable Lagrangians for Massive
Yang-Mills Fields}, {\it Nucl. Phys. \bf 35}, 167 (1967).}

\REF\gmlow{Gell-Mann, M. and Low, F., {\sl   Quantum Electrodynamics at
Small Distances}, {\it Phys. Rev. \bf 95},  1300  (1954).}

\REF\stuck{Stueckelberg, E. and Petermann, A., {\sl La Normalisation des
Constantes dans
la Theorie des Quanta}, {\it Helv. Phys. Acta
\bf 26},  499  (1953).}

\REF\calsym{Callan, C.G., {\sl Broken Scale Invariance in Scalar Field
Theory},{\it Phys. Rev \bf  D 2},1541 (1970);
Symansik, K., {\sl  Small Distance Behavior in Field Theory and Power
Counting}, {\it Comm. Math. Phys. \bf 18}, 227 (1970).}

\REF\wilren{Wilson, K., {\sl   Renormalization Group and Strong Interactions},
{\it Phys. Rev. \bf D3}, 1818  (1971).}

 \REF\light{Jackiw, R. Van Royen, R. and West, G.,
{\sl Measuring Light-Cone Singularities}, {\it Phys. Rev. \bf D2}, 2473
(1970);
Frishman, Y., {\sl Operator Products at Almost Light Like Distances}, {\it
Ann. Phys. \bf 66}, 373 (1971);
Leutwyler, H. and Stern, J., {\sl Singularities of Current Commutators on the
Light Cone}, {\it Nucl. Phys. \bf B20},77 (1970);
Gross, D. J., {\sl Light Cone Structure of Current Commutators in the
Gluon-Quark Model},
{\it Phys. Rev. \bf D4}, 1059 (1971).}

\REF\calsymt{Callan, C.G., {\sl  Broken Scale Invariance and Asymptotic
Behavior},
{\it Phys. Rev \bf  D 5},3202 (1972);
Symansik, K., {\sl Small-Distance-Behavior Analysis and Wilson Expansions},
{\it Comm. Math. Phys. \bf 23}, 49 (1971);
Christ, N., Hasslacher, B. and Mueller, A.,
{\sl Light-Cone Behavior of Perturbation Theory},
{\it Phys. Rev \bf  D6},3543 (1972).}

\REF\kowil{Wilson, K. and Kogut, J., {\sl The Renormalization Group and the
$\epsilon$ expansion.   }{\it Phys. Rep. \bf 31}, 792 (1973).}

\REF\Par{Parisi, G., {\sl Deep Inelastic Scattering in a Field Theory with
Computable
Large-Momenta Behavior}, {\it Lett. al Nuovo Cimento \bf 7}, 84 (1973).}

\REF\grocall{Callan, C. G.and Gross, D. J., {\sl  Bjorken Scaling in Quantum
Field Theory}, {\it Phys. Rev. \bf D8},  4383 (1973).}

\REF\grocol{ Coleman, S. and Gross, D. J., {\sl Price of Asymptotic Freedom},
{\it Phys. Rev. Lett. \bf 31}, 851 (1973).}

\REF\zee{Zee, A.,  {\sl Study of the Renormalization Group for Small Coupling
Constants},    {\it Phys. Rev. \bf D7},  3630  (1973).}

\REF\growil{Gross, D. J. and Wilczek, F., {\sl   Ultraviolet Behavior of
Non-Abelian Gauge Theories}, {\it Phys. Rev. Lett. \bf 30}, 1343 (1973).}

\REF\pol{Politzer, H.D., {\sl   Reliable Perturbative Results for Strong
Interactions?}, {\it Phys. Rev. Lett. \bf 30}, 1346 (1973).}

\REF\Niel{Nielsen, N.K., {\sl   Asymptotic Freedom as a Spin Effect},
{\it Am. J. Phys. \bf 49}, 1171 (1981).}

\REF\bfgm{Bardeen, W.A., Fritzsch, H. and Gell-Mann, M., {\sl Light-Cone
Current
Algebra, $\pi^0 $ Decay, and $e^+e^-$ Annihilation}, in {\it Scale and
Conformal
Symmetry in Hadron Physics}, ed. R. Gatto (Wiley \& Sons, 1973).}

\REF\wein{Weinberg, S., {\sl   Non-Abelian Gauge Theories of the Strong
Interactions}, {\it Phys. Rev. Lett. \bf 31}, 494 (1973).}

\REF\fgml{Fritzsch, H., Gell-Mann, M. and Leutwyler, H., {\sl Advantages
of the Color Octet Gluon Picture}, {\it Physics Lett. \bf 47B}, 368 (1973).}

\REF\fglm{Fritzsch, H. and Gell-Mann, M., {\sl   Current Algebra: Quarks and
What Else?}, in {\it  Proceedings of the XVI International Conference on High
Energy Physics}, Vol. 2, 164 (1972).}

\REF\instant{ Belavin, A., Polyakov, A., Schwartz, A.  and Tyupkin, Yu.,
{\sl Psuedoparticle Solutions of the Yang-Mills Equations},
{\it Phys. Lett. \bf 59B}, 85 (1975).}

\REF\thooftinst{'t Hooft, G., {\sl Computation of the Quantum Effects due to
a Four-Dimensional Psuedoparticle}, {\it Phys. Rev. \bf D14}, 3432 (1976).}

\REF\thetaa{Callan, C., Dashen, R. and Gross, D, {\sl The Structure of the
Vacuum}, {\it Phys. Lett.} (1976); Jackiw, R. and Rebbi, C. {\sl
Vacuum Periodicity in a Yang-Mills Quantum Theory}, {\it Phys. Rev. Lett.
\bf 37}, 172 (1976).}

\REF\growill{Gross, D. J. and Wilczek, F., {\sl    Asymptotically Free Gauge
Theories.I}, {\it Phys. Rev. \bf D8}, 3633 (1973).}

\REF\growilll{ Gross, D. J. and Wilczek, F., {\sl   Asymptotically Free Gauge
Theories.II}, {\it Phys. Rev. \bf D9}, 980 (1974).}

\REF\lipkin{ Lipkin, H. {\sl Triality, Exotics and the Dynamical Basis
of the Quark Model}, {\it Phys. Lett. Vol. \bf 45B}, 267 (1973).}

\REF\pagels{Marciano, B. and Pagels, H., {\sl Quantum Chromodynamics}, {\it
Phys. Rep.
\bf C 36  }, 137
(1978).}

\REF\wilcon{Wilson, K., {\sl   Confinement of Quarks}, {\it Phys. Rev. \bf
D10}, 2445 (1974).}

\REF\Creuz{ Creuz, M., {\sl Confinement and the Critical
Dimensionality of Space-Time,} {\it Phys. Rev. Lett. \bf 43}, 553     (1968).}

\REF\gronev{Gross, D. J. and Neveu, A., {\sl  Dynamical Symmetry Breaking
in an Asymptotically Free Theory}, {\it Phys. Rev.  \bf  D10}, 3235 (1974).}

\REF\thoot{t' Hooft, G., {\sl A Planar Diagram Theory for Strong
Interactions}, {\it  Nucl. Phys. \bf B72}, 461 (1974).}

\REF\casher{A. Casher, J. Kogut and L. Susskind, {\sl Vacuum Polarization and
the
Quark-Parton puzzle}, {\it Phys. Rev. Lett. \bf 31}, 792 (1973); {\sl Vacuum
Polarization and the Absence of Free Quarks}, {\it Phys. Rev. \bf D 10}, 732
(1974).}

\REF\cas{Caswell, W., {\sl Asymptotic Behavior of Non-Abelian Gauge Theories
to Two-Loop Order},  {\it Phys. Rev. Lett. \bf 33}, 244 (1974).}

\REF\Jones{Jones, D. {\sl Two-Loop Diagrams in Yang-Mills Theory},  {\it Nucl.
Phys. \bf B75}, 531,   (1974).}

\REF\appelt{Appelquist, T. and Georgi, H., {\sl $e^+e^-$ Annihilation in Gauge
Theories of Strong Interactions},
{\it Phys. Rev. \bf D8}, 4000 (1973).}

\REF\Zet{Zee, A., {\sl Electron-Positron Annihilation in
Stagnant Field Theories}, {\it Phys. Rev. \bf D8}, 4038, (1973).}

\REF\MaryK{ Gaillard, M.K. and Lee, B.W.{\sl $\Delta I ={1\over 2}$ Rule for
Non-Leptonic Decays in Asymptotically Free Field Theories},
{\it Phys. Rev. Lett \bf 33}, 108 (1974).}

\REF\Alt{ Altarelli, G. and Maiani, L., {\sl Octet Enhancement of Non-Leptonic
Weak
Interactions in Asymptotically Free Gauge Theories}, {\it Phys. Lett. \bf 52B},
351 (1974).}

\REF\scaletest{Gross, D. J., {\sl   How to Test Scaling in Asymptotically Free
Theories}, {\it Phys. Rev. Lett. \bf 32}, 1071 (1974).}

\REF\bim{Bouchiat,C., Iliopoulpos, J. and Meyer, Ph., {\sl An
Anomaly Free Version of Weinberg's Model},
{\it Phys. Lett. \bf 38B}, 519 (1972).}

\REF\grojak{Gross, D.J. and Jackiw, R., {\sl The Effect of Anomalies on
Quasi-Renormalizable Theories},
{\it Phys. Rev. \bf D6}, 477 (1972).}

\REF\GaLeRos{ Gaillard, M.K., Lee, B.W. and Rosner, J.L., {\sl Search for
Charm},
{\it Rev. Mod. Phys. \bf 47}, 277 (1975)}

\REF\anom{ The discovery and application of axial anomalies had a big impact
on  the development of the standard model. The original observation was
due to S. Adler, {\it Phys. Rev. \bf 177}, 2426 (1969);  and independently, J.
Bell
and R. Jackiw, {\it Nuovo Cimento \bf 60A}, 47 (1969). They
discovered that the axial vector current was not conserved in gauge theories,
as one would naively expect, but rather proportional to
$F^{\mu \nu}\tilde F_{\mu \nu}  $. This explained the invalidity of the
naive low energy theorem  which predicted a vanishing $\pi^0$ decay
rate to two photons in the chiral limit. The demonstration of Adler, and
S. Adler and W. Bardeen, {\it Phys. Rev. \bf 182}, 1517 (1969),
that the coefficient of the anomaly was given {\em exactly}, by lowest order
perturbation theory (the triangle graph) meant that
the pion decay constant could yield
information about the sum of the squares of the charges
of the elementary constituents of the hadrons. This was used as one of
the arguments for color \refmark{\bfgm}. The proof of non-renormalization
of the anomaly was generalized to
non-Abelian gauge theories by Bardeen. When 't Hooft's demonstration
of the renormalizability
of the electro-weak theory \refmark{\thooft} appeared many of us worried
whether the axial
anomalies would invalidate the proof. The argument    that  anomalies
do spoil renormalizability, and
the suggestion that the anomalies would cancel between quarks and leptons
if there was a charmed quark, gave further impetus to the the idea
of charm \refmark{\grojak,\bim}.
Later, the discovery of instantons \refmark{\instant},  produced
a convincing argument \refmark{\thooftinst}~as to
the lact of a chiral U(1) symmetry of hadrons, which was
in conflict with the observed spectrum of psuedoscalar mesons,
and revealed the $\theta$ parameter  of QCD \refmark{\thetaa}.}

\REF\appelt{Appelquist, T. and Politzer, H. D., {\sl Heavy Quarks and $e^+ e^-$
Annihilation},
{\it Phys. Rev. Lett. \bf 34}, 43 (1975).}

\REF\georglas{Georgi, H. and Glashow, S., {\sl  Unity of All Elementary
Particle Forces}, {\it Phys. Rev. Lett. \bf 32 }, 438  (1974).}

\REF\Pati{Pati, J.C. and Salam, A., {\sl  Unified Lepton-Hadron Symmetry and a
Gauge Theory of the Basic  Interactions}, {\it Phys. Rev.   \bf D 8 }, 1240
(1973).}

\REF\Quinn{Georgi, H., Quinn, H.R. and Weinberg, S., {\sl Hierarchy of
Interactions in Unified Gauge Theories}, {\it Phys. Rev. Lett. \bf 33}, 451
(1974).}

\Pubnum{PUPT 1329}
\date={June, 1992}
\titlepage
\vsize=23.5 true cm
\hsize=15.4 true cm
\title{ {\bf ASYMPTOTIC   FREEDOM AND THE EMERGENCE OF QCD}
\footnote*{Talk delivered at the Third International
Symposium on the History of Particle Physics,  June 26, 1992}}
\author{DAVID J. GROSS\footnote\dagger{Supported by NSF Grant PHY90-21984} }
\address{\JHL}
\vskip 1truein

\abstract

        I describe  our understanding of the strong interactions
at the end of the 1960's and the impact of the experiments on deep inelastic
scattering. I  recall  the steps that lead from the attempts to understand
these
experiments to the discovery of asymptotic freedom in 1973 and the subsequent
rapid emergence, development  and acceptance of
the non-Abelian gauge theory of color (QCD) as the theory of the strong
interactions.

\chapter{\bf  INTRODUCTION}

        This is a  marvelous occasion on which we   meet  to explore the
developments that led to  the construction of the Standard Model of elementary
particle physics. The Standard Model is surely one of the major intellectual
achievements of the twentieth century. In the late 1960's and early 1970's,
decades of path breaking experiments culminated in the emergence of a
comprehensive theory of particle physics. This theory identifies the basic
fundamental constituents of matter and describes all the forces of nature
relevant
at accessible energies-- the strong, weak and electromagnetic interactions.

                We are here to recount the history of this development. My task
is to discuss the history of the emergence of QCD, with particular emphasis on
the discovery of asymptotic freedom. Science progresses
in a much more muddled fashion than is often pictured in history
books. This is especially true of theoretical physics, partly because history
is written by the victorious. Consequently, historians of science often
ignore the many alternate paths that people wandered down, the
many false clues they   followed,  the many misconceptions they had. These
alternate points of view  are less clearly developed  than the final theories,
harder
to understand  and easier to forget, especially as these are viewed
years later, when it all  really does make sense. Thus reading history one
rarely
gets the feeling of the true nature of scientific development, in which the
element
of farce is as great as the element of triumph.

        The emergence of QCD  is a wonderful  example
of the evolution from farce to triumph.
During a very short period,  a transition occurred from
experimental discovery and   theoretical confusion to theoretical  triumph
and experimental confirmation.  We  were lucky to have been young then,
when we could stroll along the newly opened  beaches
and pick up the many  beautiful shells that  experiment had revealed. In trying
to
relate this story, one  must be wary of the danger of the personal bias that
occurs
as one looks back in time. It is not totally possible to avoid this.
Inevitably, one is
fairer to oneself than to others, but one can try. In any case the purpose of
this
meeting, I gather, is to provide raw material for the historians. One can take
consolation from  Emerson, who  said that {\sl    \lq\lq There is properly no
history; only biography."}

\chapter{ \bf THE THEORETICAL SCENE}

        I would like first to describe the  scene in theoretical particle
physics, as I saw it in the early 1960's at Berkeley,  when I started as  a
graduate
student.  The state of particle physics was then almost the complete opposite
of
today. It was a  period  of experimental supremacy and theoretical impotence.
The construction and utilization of major accelerators were proceeding at full
steam. Experimental discoveries and surprises appeared every few months.
There was hardly any theory to speak of. The emphasis was on phenomenology,
and  there were only small islands of theoretical advances here and there.
Field
theory was in disgrace; S-Matrix theory was in full bloom. Symmetries were all
the rage.

The field was divided into the study of the weak and the strong interactions.
In
the case of the weak interactions, there was a rather successful
phenomenological theory, but not much new data.  The strong interactions were
where the experimental and theoretical action was, particularly at Berkeley.
They
were regarded as especially unfathomable.
The prevalent feeling was  that it would take a very long time to
understand the strong interactions and that it would require revolutionary
concepts.   For a young graduate student this was clearly the major challenge.

The feeling at the time was well expressed  by Lev Landau
in his last paper, called \lq\lq Fundamental Problems,'' which appeared in a
memorial volume to Wolfgang Pauli in 1959 \refmark{\lan}. In this paper  he
argued that
quantum field theory had been nullified by the discovery of the zero charge
problem. He said:

{\sl   \lq\lq It is well known that theoretical physics is at present almost
helpless in dealing with the problem of strong interactions.... By now the
nullification of the theory is tacitly accepted even by theoretical
physicists who profess to dispute it. This is evident from the almost
complete disappearance of papers on meson theory and particularly from
Dyson's assertion that the correct theory will not be found in the next
hundred years.''}

Let us explore  the theoretical  milieu at this time.
\smallskip
\section{\bf Quantum Field Theory}
\smallskip

Quantum field theory was originally developed for the treatment of
electrodynamics almost immediately after the completion of quantum mechanics
and
the discovery of the Dirac equation. It seemed to be   the natural tool for
describing the dynamics of elementary particles. The application of quantum
field
theory    had important early  success. Fermi formulated a powerful and
accurate
phenomenological  theory of beta decay, which was to serve as a framework for
exploring the weak interactions for three decades. Yukawa proposed a field
theory to describe the nuclear force and predicted the existence of  heavy
mesons, which were soon discovered. On the other hand, the theory was
confronted from the beginning with severe  difficulties. These included the
infinities that appeared as soon as one went beyond lowest order perturbation
theory, as well as the   lack of  any non-perturbative understanding of
dynamics.
By the 1950's the suspicion of field theory had  deepened  to the point that a
powerful dogma emerged--that field theory  was fundamentally wrong, especially
in its application to the strong interactions.

        The renormalization procedure, developed by Richard Feynman, Julian
Schwinger,
Sin-itiro Tomanaga and Freeman Dyson, was spectacularly successful in Quantum
Electrodynamics. However, the physical meaning  of renormalization was not
truly understood. The feeling of most was that renormalization was a trick.
This was especially the case for the pioneering inventors of quantum field
theory (for example Dirac and Wigner). They were prepared at the first drop of
an
infinity to renounce their belief in quantum field theory and to brace for the
next  revolution. However it was also the feeling of the younger leaders of the
field,
who had laid the foundations of perturbative quantum field theory and
renormalization in the late \rq 40's. The prevalent feeling was that
renormalization
simply swept the infinities under the rug, but that they were still there and
rendered the notion of local fields meaningless. To quote Feynman, speaking at
the 1961 Solvay conference\refmark{\fey}, {\sl    \lq\lq I still hold to this
belief and
do not subscribe to the philosophy of renormalization."}

Field theory was almost totally perturbative at that time. The nonperturbative
techniques that had been tried in the 1950's had all failed.
The path integral, developed by Feynman in the late 1940's,
which later proved so valuable for a nonperturbative formulation of quantum
field theory as well as a tool for semiclassical expansions and numerical
approximations, was almost completely forgotten. In a sense the   Feynman
rules
were  too successful. They were an immensely useful,  picturesque and intuitive
way of performing perturbation theory. However these alluring qualities also
convinced
many  that all that was needed from field theory were these rules. They
diverted
attention from the non-perturbative dynamical issues facing field theory.
In my first course on quantum field theory at Berkeley in 1965, I was taught
that
{\em Field Theory =  Feynman Rules.}

No examples were known  of four dimensional field theories that
one could handle non-perturbatively.  Indeed, except for free field theory and
a few examples of soluble two dimensional field theories, there were no models
that could serve as practice grounds for developing field theoretic tools and
intuition. The more mathematically inclined warned us that
most naive manipulations in quantum field  theory  were unjustified and that
the whole structure was potentially unsound. They initiated a program of
axiomatic analysis that promised to inform us in a decade or two
whether quantum field theory made sense.

 In the United States however I think the main reason for the abandonment of
field theory was simply that one could not calculate.  American physicists are
inveterate  pragmatists. Quantum field theory had not proved to be a useful
tool
with which to  make contact with the explosion of  experimental discoveries.
The early attempts in the 1950's to construct field theories of the strong
interactions were total failures. In hindsight this was not surprising since a
field theory of the strong interactions faced two enormous problems.
First, which fields to use? Following Yukawa,  the first attempts  employed
pion
 and nucleon fields.  Soon, with the rapid proliferation of particles, it
became
 evident  that nothing was special about the nucleon or the pion. All the
hadrons, the strange baryons and mesons as well as the higher spin recurrences
of these, appeared to be equally fundamental.
The obvious conclusion that all hadrons were composites of more
fundamental constituents was thwarted by the fact that
no matter how hard one smashed hadrons
at each one had not been able to
liberate these hypothetical constituents.
This was not analogous to the paradigm of atoms made of nucleons and electrons
or of
nuclei composed of nucleons. The idea of permanently bound, confined,
constituents was unimaginable at the time.  Second,  since the pion-nucleon
coupling was so large, perturbative expansions were useless. All
attempts at non perturbative analysis  were unsuccessful.

        In the case of the weak interactions, the situation was somewhat
better. Here one had an adequate effective theory--the four fermion
Fermi interaction, which could be usefully employed, using perturbation
theory to lowest order, to organize and understand the
emerging experimental picture of the weak interactions. The fact that this
theory was  non-renormalizable meant  that beyond  the Born approximation
it lost all predictive value. This disease increased the  suspicion of field
theory.
Yang-Mills theory, which had appeared in the mid 1950's was not taken
seriously. Attempts to apply Yang-Mills theory to the strong interactions
focused on elevating global flavor symmetries to local gauge symmetries.
This was problematic since these symmetries were not exact.
In addition non-Abelian gauge theories apparently required
massless vector mesons--clearly not a feature of the strong interactions.

        In the Soviet Union field theory was under even heavier attack,
for somewhat different reasons.  Landau and collaborators, in the late 1950's,
studied
the high  energy behavior of
quantum electrodynamics.  They explored the relation
 between the physical electric charge  and the bare electric charge
(essentially
the electric charge   that controls the physics at energies of order the
ultraviolet
cutoff). They concluded, on the basis of their approximations,
that the physical charge  vanishes, for any value of the bare
charge as we let the ultraviolet cutoff become infinite (this is of
course necessary to achieve a Lorentz invariant theory).\refmark{\lanpom}

{\sl ``We reach the conclusion that within the limits of
formal electrodynamics a point interaction is equivalent, for any intensity
whatever, to no interaction at all.'' }

 This is the famous  problem of {\em zero charge},
a startling result that implied for Landau that {\sl  \lq\lq weak
coupling electrodynamics is a theory which is, fundamentally,
logically incomplete.''}~\refmark {\lanft}~This  problem  occurs in any non-
asymptotically-free theory. Even today, many of us  believe that
a  non-asymptotically-free theory such as QED, if taken by itself, is
inconsistent  at very high energies.  In the case of QED this is only an
academic
problem, since the trouble shows up only at enormously high energy. However in
the case of the strong interactions, it was an immediate catastrophe.  In the
Soviet
Union this was thought  to be a compelling reason why field theory was wrong.
Landau decreed that \refmark{\lan}

{\sl    \lq\lq  We are driven to the conclusion that the Hamiltonian method for
strong interaction is dead and must be buried, although of course
with deserved honor.''}

Under the influence of Landau and Pomeranchuk, a generation of physicists was
forbidden to work on field theory. One might wonder why the discovery of the
zero
charge  problem did not inspire a search for asymptotically free theories that
would be free of this disease. The answer, I think, is twofold. First, many
other
theories were explored--in each case they behaved as QED. Second, Landau and
Pomeranchuk concluded, I think, that this problem was inherent in any quantum
field  theory, that an asymptotically free theory could not exist.\footnote*{
As Kenneth
Johnson pointed out \refmark{\John}, a calculation of the charge
renormalization
of charged vector mesons was carried out by V.S. Vanyashin and  M.V. Terentev
in
1964\refmark{\Van}. They got the magnitude wrong but did
get the correct sign and
concluded that the result was absurd. They attributed this  wrong sign to the
non-renormalizability of charged vector meson theory.}
\smallskip
\section{\bf The Bootstrap}
\smallskip
The bootstrap theory rested on two principles, both more philosophical
than scientific. First, local fields were not directly measurable.
Thus they were unphysical and meaningless.
Instead, one should formulate the theory using only  observables.
The basic observables  are the S-Matrix elements   measured in scattering
experiments. Microscopic dynamics was renounced. Field theory was to be
replaced by
S-matrix theory; a theory  based on general principles, such as
unitarity and analyticity,   but with no fundamental microscopic Hamiltonian.
The basic dynamical idea was that there was a unique S-Matrix that obeyed these
principles. It could be determined without the unphysical demand of fundamental
constituents or equations of motion that was inherent in field theory. To quote
Geoffrey Chew \refmark{\smat},

{\sl \lq\lq Let me say at once that I believe the conventional association of
fields with strong interacting particles to be empty. It seems to me that no
aspect  of strong interactions has been clarified by the field concept.
Whatever
success theory has achieved in this area is based on the unitarity
of the analytically continued S-matrix plus symmetry principles.  I do not wish
to assert (as does Landau) that conventional field theory is necessarily wrong,
but only that it is sterile with respect to the strong interactions and that,
like  an old soldier, it is destined not to die but just to fade away."}

In hindsight, it is clear that the bootstrap  was born from the frustration of
 being unable to calculate anything using field theory. All models and
approximations produced conflicts with some dearly held principle.  If it was
so
difficult to construct an S-Matrix that was consistent with sacred principles
then
maybe these general principles had a unique manifestation.
The second principle of the bootstrap was that there were no elementary
particles.
The way  to
deal with the increasing number of  candidates for elementary status was to
proclaim that all were equally fundamental, all were dynamical bound states of
each other. This was called  Nuclear Democracy, and was a response to the
 proliferation of candidates for fundamental building blocks.  As Chew
stated \refmark{\chew},

  {\sl \lq\lq The notion, inherent in conventional Lagrangian field theory,
that certain particles are fundamental while others are complex,
is becoming less and less palatable for baryons and mesons
as the number of candidates for elementary status continues to increase.}"

  The bootstrap idea was immensely popular in the early 1960's, for a
variety of reasons. Superseding quantum field theory, it rested on the solid
 principles of causality and unitarity. It was real and physical.  It promised
to be very predictive, indeed to provide a unique value for all observables,
satisfying a basic desire of particle physicists to believe that the world
around us is not arbitrary and that, to quote Einstein,

  {\sl    \lq\lq  Nature is so constituted that it is possible logically to lay
down such strongly determined laws that within these laws only rationally,
completely determined constants occur, not ones whose
numerical value could be changed without destroying the theory."}

        The bootstrap promised that this hope would be realized already in the
theory of  the strong interactions. This is of course false. We now know that
there are an infinite number of consistent  S-Matrices  that satisfy
all the sacred principles.  One can take any non-Abelian gauge theory, with any
gauge group, and many sets of fermions (as long as there are not too many to
destroy asymptotic freedom.)   The hope for uniqueness must wait for a higher
level of unification.

In Berkeley, as in the Soviet Union, S-Matrix theory was supreme, and a
generation of young theorists was raised ignorant of field theory.  Even on the
calmer East Coast S-Matrix theory swept the field.
For example, I quote Marvin Goldberger  who said \refmark{\Gold},

 {\sl \lq\lq My own feeling is that we have learned a great deal from field
theory... that I am quite happy to discard it as an old, but rather friendly,
mistress who I would be willing to recognize on the street if I should
encounter
her again. From a philosophical point of view and certainly from a practical
one the S-matrix approach at the moment seems to me by far the most
attractive."}

S-Matrix theory had some notable successes, the early application of dispersion
relations and  the development of Regge pole  theory.  However, there were
drawbacks to a theory that was based on the principle that there was no theory,
at  least in the traditional sense.  As Francis Low said \refmark{\loww},

   {\sl     \lq\lq  The distinction between S-Matrix theory and field theory
is,
 on the one hand, between a set of equations that are not formulated, and on
the
 other hand between a set of equations that are formulated if you knew what
they
 were and for which you do not know whether there is a solution or not."}

Nonetheless, until 1973 it was not thought proper to use field
theory without apologies. For example as late as the NAL conference of 1972,
Murray Gell-Mann ended his talk on quarks with the summary \refmark{\nal},

  {\sl  \lq\lq Let us end by emphasizing our main point,
that it may well be possible to construct an explicit theory of hadrons,
based on quarks and some kind
of glue, treated as fictitious, but with enough physical properties abstracted
and applied to real hadrons to constitute a complete theory. Since the entities
we start with are fictitious, there is no need for any conflict with
the bootstrap or conventional dual parton point of view.}"

\smallskip

 \section{\bf Symmetries}

 If dynamics was impossible, one could at least explore the symmetries of the
strong
interactions. The biggest advance of the early 1960's
was the discovery of an approximate
symmetry of hadrons, $SU(3)$, by Gell-Mann and Yuval Neeman, and then the
beginning
of the understanding of spontaneously broken chiral symmetry. Since the
relevant  degrees of freedom, especially color,  were totally hidden from view
due to confinement, the emphasis was on flavor, which was directly observable.
This emphasis was  enhanced because of the success of $SU(3)$. Nowadays
we realize that $SU(3)$ is  an accidental symmetry, which arises  simply
because
a few quarks (the up, down and strange quarks) are relatively light compared to
the scale of the strong interactions. At the time it  was regarded as a deep
symmetry
of the strong interactions, and many attempts were made to generalize it
and use it as a springboard for a theory of hadrons.

  The most successful attempt was Gell-Mann's algebra of  currents
\refmark{\veal}.
In an  important and beautiful paper, he outlined a program
for abstracting relations from a field theory, keeping the ones that might
be generally true and then throwing the field theory away \refmark{\veal},

{\sl  \lq\lq In order to obtain such relations that we conjecture to be true,
we use the method of abstraction from a Lagrangian field theory model. In other
words, we construct a mathematical theory of the strongly interacting
particles, which may or may not have anything to do with reality, find
suitable algebraic relations that hold in the model, postulate their validity,
and then throw away the model. We may compare this process to a method
sometimes employed in French cuisine: a piece of pheasant meat is cooked
between two slices of veal, which are then discarded."}

 This paper made quite an impression,
especially  on impoverished graduate students like me,
who could only dream of eating such a meal. It was a marvelous approach. It
gave
one  the freedom to play with the forbidden fruit of field theory, abstract
what one
wanted from it, all without having to believe in the theory.
The only problem was that it was not clear what principle
determined what to  abstract?

The other problem with this approach was that it diverted
attention from  dynamical issues.
The most dramatic example of this is  Gell-Mann and George Zweig's hypothesis
of
quarks \refmark{\quarks},
the most important consequence of the discovery of SU(3).
The fact was that hadrons looked as if
they were composed of (colored) quarks\footnote*{ Color was introduced by
O.W.  Greenberg \refmark{\green} and M.Y. Han
and Yoichiro Nambu \refmark{\hannambu} to explain
the strange statistics of non-relativistic quark model hadronic bound states.},
whose masses (either the current quark masses or the constituent quark masses)
were quite small. Yet quarks had not been seen, even when
energies were achieved that were ten times the threshold for their production.
This was not analogous to atoms  made of nuclei and electrons  or to nuclei
made of
nucleons. The non-relativistic quark model simply did not make sense.
Therefore quarks were fictitious, mathematical devices.

        It was a pity that particle theorists at that time, for the most part,
totally ignored condensed matter physics. There were of course notable
exceptions  such as Nambu, and the last of the true universalists, Landau, who
unfortunately was incapacitated at an early stage.
This attitude was largely a product of arrogance. Particle physics was much
more fundamental  and basic than the solid state physics that studied
collections
of many atoms, whose basic laws of interaction were well understood. Thus
particle physicists thought that they had little to learn
from \lq\lq dirt physics" (or \lq\lq squalid state physics").
This attitude was unfortunate. We would have profited much from a deeper study
of superconductivity--the preeminent advance in condensed matter physics in
this
period. Not only the insight it gave, stressed by Philip Anderson, into broken
gauge
symmetry-- but also of  the possibility of confinement.
The Meissner effect that occurs in the superconducting state is a very good,
dual (under interchange of electric and magnetic fields) analog of confinement.
Indeed if magnetic monopoles existed, they
would form, in the superconducting state, magnetically neutral bound
states that would be quite analogous to hadrons.  This idea was not explored by
condensed matter  physicists either, perhaps since   monopoles had not been
found. The situation would have been different if monopoles had existed to
provide a live example of confinement.

        This attitude   towards quarks persisted until 1973 and beyond. Quarks
clearly did not exist as real particles, therefore they were fictitious devices
(see Gell-Mann above). One might \lq\lq abstract" properties of quarks from
some
 model, but one was not allowed to believe in their reality  or to take
the models too seriously.

        For many this smelled fishy. I remember very well Steven
  Weinberg's reaction to the sum rules Curtis Callan
and I had derived using the quark-gluon
model. I  described my work on  deep inelastic scattering sum rules to
Weinberg at a Junior Fellows dinner at Harvard. I needed him to write a
letter of recommendation to Princeton, so I was a little nervous. I explained
how the small longitudinal cross section observed at SLAC could be interpreted,
on  the basis of our sum rule   as evidence for quarks. Weinberg was emphatic
that this was of no interest since he did not believe anything about quarks. I
was  somewhat  shattered.

 \section{\bf Experiment}

This was a period of great  experimental excitement.  However, I would like to
discuss  an interesting phenomenon, in which theorists and experimentalists
reinforced each other's conviction that the secret of the strong interactions
lay in
the high energy behavior of scattering amplitudes at low momentum transfer.
Early scattering experiments concentrated, for obvious reasons, on the events
that had the largest
rates. In the case of the strong interactions,
this meant searching  for  resonant
bumps or probing near forward scattering,
where  the cross section was largest.
 It was not at all realized by theorists that the secret of hadronic dynamics
could  be revealed by experiments at large momentum transfer that probed the
short distance structure of hadrons. Instead, prompted by the regularities that
were discovered at low momentum transfer, theorists  developed an explanation
based on the  theory of Regge poles. This was the only strong interaction
dynamics that was understood, for which there was a real  theory. Therefore
theorists concluded that Regge behavior must be very important and forward
scattering experiments were deemed to be  the major tool of discovery.
Regge theory was soon incorporated into the bootstrap program as a boundary
condition.   In response to this theoretical enthusiasm,  the  interest of
experimentalists in forward scattering was enhanced. Opportunities to probe the
less  easily accessible domains of large momentum transfer were ignored. Only
much later,
after the impact of  the deep inelastic scattering experiments  that had been
ridiculed by many as unpromising, was it understood that the
most  informative experiments were those at
large momentum  transfers that probe  short or light-like distances.

        It used to be the case that when a new accelerator was initiated one of
the first and most important experiments to be  performed was the measurement
of
the total p-p cross section. Nowadays, this experiment is regarded with little
interest,  even though the explanation of Regge behavior remains an
interesting,
unsolved and complicated problem for QCD.  Ironically, one of  the principal
justifications for this experiment today is simply to  calibrate  the
luminosity of the
machine.
\smallskip

\chapter {\bf   My Road to Asymptotic Freedom}
 \section{\bf From N/D to QCD}

I was a graduate student at Berkeley at the height of the bootstrap
and S-Matrix theory. My Ph.D. thesis was written under the supervision of Geoff
Chew, the main guru of the bootstrap, on multi-body ${N\over D}$ equations.  I
can remember the precise moment at which I was disillusioned with the bootstrap
program. This was at the  1966 Rochester meeting, held at Berkeley.
Francis Low, in the session following his talk, remarked
that the bootstrap  was less of a theory than a  tautology \refmark{\low},

{ \sl \lq\lq  I believe that when you find that the particles that are there in
S-Matrix theory, with crossing matrices and all the formalism, satisfy all
these
conditions, all you are  doing is showing that the S matrix is consistent with
the world the way it is; that is the particles have put themselves there in
such
a way that it works out, but you have not necessarily explained that they are
there."}

For example, the then popular finite energy sum rules (whereby
one derived
relations for measurable quantities by saturating dispersion relations with a
finite
number of resonance poles on the one hand and relating these to the assumed
Regge asymptotic behavior on the other) were not so much  predictive equations,
but merely checks of axioms (analyticity, unitarity) using models and fits of
experimental data.

I was very impressed with this remark and longed to find a more powerful
dynamical scheme. This was the heyday of current algebra, and
the air was buzzing with marvelous results.  I was very impressed by
the fact that one could  assume a certain structure of current
commutators and derive measurable results. The most dramatic of these
was the Adler-Weisberger relation that had just appeared\refmark{\adwe}.
Clearly the properties of these currents placed strong restrictions on hadronic
dynamics.

The most popular scheme then was current algebra.
Gell-Mann and Roger Dashen  were trying to use the commutators of certain
components
of the currents as a basis for strong interaction dynamics.
After a while I concluded that this approach  was also tautological---all it
did was
test the validity of the symmetries of the strong interactions.
This was apparent for vector $SU(3)$. However it was also true  of chiral
$SU(3)$,
especially as the current algebra sum rules were
interpreted, by Weinberg and others, as low energy theorems for Goldstone
bosons. This scheme could not be a basis for  a complete dynamical theory.

I  studied the less understood properties of the algebra of local current
densities.
These were model dependent; but that was fine, they therefore might contain
dynamical information that went beyond statements of global symmetry.
Furthermore, as was soon realized, one could check ones' assumptions about
the structure of local current algebra by deriving sum rules that could be
tested in
deep inelastic lepton-hadron scattering experiments. James Bjorken's 1967 paper
\refmark{\bj,\bjj}, on the application of  $U(6)\times U(6)$, particularly
influenced
me.

        In the spring of 1968 Curtis Callan and I proposed a sum rule to test
the then popular \lq \lq Sugawara model," a dynamical model
of local currents, in
which the energy momentum tensor was expressed as a product
of currents\refmark{\sug }.
The hope was that the algebraic properties of the currents
and the expression for the Hamiltonian in terms of these
would be enough to have a complete theory.
(This idea actually works
in the now very popular two dimensional conformal field theories).
Our goal was slightly
more modest--to test the hypothesis by
exploiting the fact that in this theory the operator
product expansion of the currents contained the
energy momentum tensor with a known coefficient.
Thus we could derive a sum rule
for the structure functions\refmark{\grocalsug}
that could be measured in deep-inelastic  electron-proton scattering.

In the fall of 1968, Bjorken noted that this sum rule, as well as dimensional
arguments, would suggest the scaling of deep inelastic scattering cross
sections
\refmark{\bjjj}. This prediction was shortly confirmed by the new experiments
at
SLAC, which were to play such an important role in elucidating the structure of
hadrons \refmark{\bloom}. Shortly thereafter Callan and I discovered that by
measuring the ratio, $R={\sigma_L\over \sigma_T}$, (where $\sigma_L$
($\sigma_T$)
is the cross section for the scattering of longitudinal  or transverse
polarized
virtual photons), one could determine  the spin of the  charged
constituents of the
nucleon \refmark{\grcal}.
We evaluated the  moments of the deep-inelastic structure
functions in terms of the equal time commutators of the electromagnetic using
specific
models for these--the {\em algebra of fields} in which the current was
proportional to a spin-one field on the one hand, and  the quark-gluon model on
the other.
In this  popular model quarks interacted through an Abelian gauge
field (which could, of course, be massive)  coupled to baryon number.
The gauge dynamics of the gluon had never been explored,
and I do not think that the model had   been used to calculate anything
until then.
We discovered that $R$ depended crucially on the spin of the constituents. If
the constituents had spin zero or one,
then $\sigma_T=0$, but if they had  spin-$ {1 \over2}$,
then $\sigma_L=0$.
This was a rather dramatic result. The experiments quickly showed that
$\sigma_L $ was very small.

        These SLAC deep-inelastic scattering experiments had a profound impact
on me. They clearly  showed that the proton behaved, when observed over short
times, as if it was made out of point-like objects of spin one-half. In
the spring of 1969,  which  I spent at CERN, Chris Llewelynn-Smith and I
analyzed the sum rules that followed for deep-inelastic
neutrino-nucleon scattering using similar methods\refmark{\grlwsm}.
We were clearly motivated by the  experiments that were then being
performed
at CERN. We  derived a sum rule that measured the baryon
number of the charged constituents of the proton. The experiments  soon
indicated that the constituents of
the proton had  baryon number ${1 \over 3}$---in other words again they
looked like quarks. I was then totally convinced of the reality of quarks.
They had to be more than just mnemonic devices for summarizing hadronic
symmetries, as they were
then universally regarded. They had to be physical point-like constituents of
the nucleon. But how could that be? Surely strong interactions
must exist between
the quarks that  would smear out their point-like behavior.

        After the experiments at SLAC, Feynman came up with his {\em parton}
picture of deep inelastic scattering. This was a very picturesque and intuitive
way of describing
deep-inelastic scattering in terms of assumed point-like
constituents--partons \refmark{\parton}.  It complemented
the approach to deep inelastic scattering based on the operator product of
currents, and had the advantage of being extendible to other processes
\refmark{\drell}. The parton model allowed one
to make predictions with ease, ignoring the dynamical issues at
hand. I felt more
comfortable with the approach based on assuming properties of current products
at short distances. I felt
somewhat uneasy about the extensions of the parton model
to processes that were not truly dominated by short distance singularities.

At CERN I studied, with Julius Wess, the consequences of exact
scale and conformal invariance \refmark{\grws}. However, I soon realized that
in
a field  theoretic context only  a free, non-interacting
theory could produce exact scaling. This became very clear to me in 1970, when
I
came  to  Princeton, where my colleague Curtis Callan (and Kurt Symansik) had
rediscovered the renormalization group equations \refmark{\gmlow}, which they
presented as a consequence of a scale invariance {\em anomaly}
\refmark{\calsym}.
Their work made it abundantly clear that
once  one introduced interactions into the theory,
scaling, as well as my beloved sum rules, went
down the tube. Yet the
experiments indicated that scaling was in fine shape. But one could hardly turn
off the interactions between the quarks, or make them very weak, since then one
would expect hadrons to
break up easily into their quark constituents. Why then had no one ever
observed free quarks? This paradox and the search for an explanation
of scaling were to preoccupy me for the following four years.
\smallskip
\section{\bf How to Explain Scaling}
\smallskip
        About the same time that all this was happening, string theory was
discovered, in one of the most bizarre turn of events in the history of
physics.
In 1968 Gabrielle Veneziano came up with a remarkably simple formula that
summarized many
features of hadronic scattering. It had Regge asymptotic behavior in one
channel and
narrow
resonance saturation in the other\refmark{\Ven}. This formula was soon
generalized to multi-particle S-Matrix amplitudes and attracted much attention.
The dual resonance model was born, the last serious attempt to implement the
bootstrap. It was only truly understood as a theory of quantized strings in
1972.
I worked on this theory for two
years, first at CERN and then at Princeton with John Schwarz and Andre Neveu.
At first I felt that this model, which captured many of the
features of hadronic scattering, might provide the long sought alternative to
a field theory of the strong interactions.
However  by 1971 I realized that there was no way that this model
could explain scaling, and I felt strongly that scaling was the paramount
feature
of the strong interactions. In fact the dual resonance model
lead to incredibly soft behavior at  large momentum transfer, quite the
opposite of the hard
scaling observed. Furthermore, it was clear that it   required for consistency
many  features that were totally unrealistic for the strong
interactions--massless
vector and tensor particles. These features later became the motivation for the
hope
that string theory may provide a
comprehensive and unified theory of all the forces of nature. This hope
remains
strong.
However the relevant energy scale is not 1 Gev  but  rather $10^{19}$Gev !

 The data on deep inelastic scattering were getting better. No violations of
scaling were observed, and the free-field-theory  sum rules worked.
I remember well the 1970 Kiev conference on high
energy physics. There I met Sasha Polyakov and Sasha Migdal, uninvited,  but
already impressive participants at the meeting. Polyakov,  Migdal and I had
long discussions about deep inelastic scattering. Polyakov knew all about the
renormalization group and explained to me that naive scaling can not be right.
Because of
renormalization the dimensions of operators change with the scale of the
physics being probed. Not only that, dimensionless couplings also change with
scale.
They  approach at
small distances fixed point values that are generically those of a
strongly coupled theory, resulting in large anomalous scaling behavior
quite different from free field theory behavior. I retorted that the
experiments
showed otherwise. He responded that this behavior contradicts field theory. We
departed; he convinced, as many were, that experiments at higher
energies would change, I that the theory would have to be changed.

The view that the scaling observed at SLAC was not a truly asymptotic
phenomenon was rather widespread. The fact that scaling set in at  rather
low  momentum transfers, \lq\lq precocious scaling,"
reinforced this view. Thus
the  cognoscenti of the renormalization group(Wilson, Polyakov,   and others)
believed that the non-canonical scaling indicative of a
non-trivial fixed point of the renormalization group
would appear at higher energies.

Much happened during the next two years. Gerhard 't Hooft's spectacular
work \refmark{\thooft}
on the renormalizability of Yang-Mills theory, reintroduced
non-Abelian gauge theories to the community.  The electroweak theory of Sheldon
Glashow, Weinberg and Abdus Salam was revived. Field theory became popular
again,
at least in application to the weak interactions.
The path integral reemerged from obscurity.

Kenneth Wilson's development of the operator product
expansion provided a  tool that could be applied to the analysis of deep
inelastic
scattering. Most important from my point of view was
the revival of the renormalization group by
Wilson
\refmark{\wilren}. The renormalization group stems from the fundamental work
of Gell-Mann and Low\refmark{\gmlow} and E. Stueckelberg and
A. Petermann\refmark{\stuck}.
This work was neglected for many years, partly because
it seemed to provide only
information about physics for large
space-like momenta, which are of no direct physical interest. Also, before the
discovery of
asymptotic freedom, the ultraviolet behavior
was not calculable
using perturbative methods, and there were no others.
Thus it appeared that the renormalization group provided a framework in which
one could discuss,
but not calculate, the asymptotic behavior of amplitudes in a physically
uninteresting region.
Wilson's development of the operator product expansion provided a new
tool that could be applied to the analysis of deep inelastic scattering.
The Callan-Symansik equations simplified the renormalization group
analysis, which was then applied to the Wilson
expansion\refmark{\calsymt,\grocall}.
The operator product analysis was extended to the light cone, the relevant
region for deep-inelastic scattering\refmark{\light}.
Most influential was Wilson's deep understanding of renormalization,
which he was then applying to critical behavior.
Wilson gave a series of lectures at Princeton in the spring of
1972\refmark{\kowil}. These had a
great impact on many of the participants, certainly on me.

  	So by the end of 1972, I had learned enough field theory, especially
renormalization group methods from Ken Wilson, to tackle the problem of
scaling  head on.  I decided, quite deliberately, to prove that local field
theory could not explain the experimental fact of scaling and thus was
not an appropriate framework for the description of the strong interactions.
Thus, deep inelastic scattering would finally settle the issue as to the
validity of quantum field theory.

The plan of the attack was twofold. First, I would prove that \lq\lq
ultraviolet
stability," the vanishing of the effective coupling at short distances,  later
called asymptotic freedom, was necessary to explain scaling. Second, I would
show that there existed no asymptotically free field theories. The latter was
to
be expected. After all the paradigm of quantum field
theory-Quantum Electrodynamics (QED)- was {\em infrared stable}; in other
words, the effective charge grew larger at short distances and no one
had ever constructed a theory in which the opposite occurred.

        Charge renormalization is nothing more (certainly in the case of
QED) than vacuum polarization. The vacuum or  the ground state
of a relativistic quantum mechanical system can be thought
of as a medium of  virtual particles. In QED the vacuum
contains virtual electron-positron pairs. If a charge, $e_0$, is put in
this medium,  it polarizes it. Such a medium with virtual electric
dipoles will  screen the charge and the actual, observable, charge e,
will differ from $e_0$ as ${e_0\over \epsilon}$, where $\epsilon$ is
the dielectric constant. Now $\epsilon$ is frequency dependent (or energy or
distance  dependent). To deal with this  one can introduce the notion of an
effective coupling $e(r)$, which governs the force at a
distance $r$. As $r$ increases, there is more medium that screens, thus $e(r)$
decreases with increasing $r$, and correspondingly
increases with decreasing $r$! The $\beta$-function, which is simply minus the
derivative of ${\rm  log}[e(r)]$ with respect to ${\rm  log}(r)$, is therefore
positive.

    If the effective coupling were, contrary to QED, to decrease
at short distances, one might explain how the strong interactions turn off in
this regime and produce scaling. Indeed, one might suspect that this
is the only way to get point-like behavior at short distances. It was
well understood, due to Wilson's work and its application to deep-inelastic
scattering, that one might expect to get scaling in a quantum field theory at a
fixed point of the renormalization group. However this scaling would not have
canonical,
free-field-theory-like
behavior. Such behavior would mean that the scaling dimensions of the operators
that appear in the product of electromagnetic currents at
light-like distances had canonical, free field
dimensions. This seemed unlikely. I knew that if the fields themselves
had canonical dimensions, then for many theories this implied that the theory
was trivial,  \ie~free.  Surely this was also true if the composite operators
that
dominated the amplitudes for
deep-inelastic scattering had canonical dimensions.

By the spring of 1973, Callan and I had completed a proof of this argument,
extending
an idea of Giorgio Parisi \refmark{\Par} to all
renormalizable field theories, with the exception of non-Abelian gauge
theories.
The essential idea was to prove that the vanishing
anomalous dimensions of the composite operators, at an
assumed fixed point of the renormalization group,
implied the vanishing anomalous dimensions of the fields. This then implied
that the theory was free at this fixed point.  The conclusion was that naive
scaling
could be explained only if the assumed fixed point of the renormalization group
was at the origin of coupling space--\ie the theory must be asymptotically
free\refmark{\grocall}. Non-Abelian gauge theories were not included in the
argument since both arguments broke down for these theories. The discovery
of asymptotic freedom made this omission irrelevant.

 The second part of the argument was to show that there were no asymptotically
free theories at all. I had set up the formalism to analyze the
most general renormalizable field theory of fermions and scalars--
again excluding non-Abelian gauge theories. This was not difficult,
since to investigate asymptotic freedom
it suffices to study the behavior of the $\beta$-functions in the vicinity of
the origin of coupling constant space, \ie~in lowest order perturbation theory
(one-loop approximation).  I almost had a complete proof but was stuck on my
inability to prove a necessary inequality. I discussed the issue with Sidney
Coleman,   who
was spending the spring semester in Princeton. He came up with the missing
ingredient, and added some other crucial points --and we had a proof that
no renormalizable field theory that consisted of theories with arbitrary
Yukawa,
scalar or Abelian gauge interactions could be asymptotically free
\refmark{\grocol}.  Tony Zee had also been studying this. He too
was well aware of the advantages of an asymptotically free theory and was
searching for one. He  derived, at the same time, a partial result,
indicating the lack of asymptotic freedom in theories with $SU(N)$ invariant
Yukawa couplings.\refmark{\zee}

\smallskip
\section{\bf The Discovery of Asymptotic Freedom}
\smallskip
        Frank Wilczek started work with me in the fall of 1972. He had come
to Princeton as a mathematics student, but soon discovered that he
was really interested in particle physics. He switched to the physics
department, after taking my field theory course in
1971, and started to work with me.
My way of dealing with students, then and  now, was to involve them closely
with
my current work and  very often to work with them directly.
This was certainly the case with Frank, who functioned more as a collaborator
than a student from the beginning. I told him about my program to determine
whether quantum field theory could account for scaling. We decided that we
would calculate the $\beta$-function for Yang-Mills theory. This was  the  one
hole in the line of argument I was pursuing. It had not been filled largely
because
Yang-Mills theory still seemed strange and difficult. Few calculations beyond
the
Born approximation had ever been done. Frank was interested in this calculation
for other reasons as well. Yang-Mills theory was already in use for the
electro-weak interactions, and he was  interested in understanding how these
behaved at high energy.

  Coleman, who was visiting in Princeton, asked me at one point whether anyone
had ever calculated the $\beta$-function for Yang-Mills theory. I told him that
we
were working on this. He expressed interest because
he had asked his student, H. David Politzer, to generalize the mechanism he had
explored with  Eric Weinberg--that of dynamical symmetry
breaking of an Abelian gauge theory--	to the
non-Abelian case. An important ingredient was the knowledge of the
renormalization flow, to decide whether lowest order perturbation
theory could be a reliable guide to the behavior of the energy functional.
Indeed, Politzer went ahead with his own calculation of the
$\beta$-function for Yang-Mills theory.

Our calculation proceeded slowly. I was involved in the other parts of my
program and there were some tough issues to resolve.  We first tried to
prove on general grounds, using spectral representations and unitarity, that
the theory could not be asymptotically free, generalizing the arguments of
Coleman and me  to this case. This did not work, so we proceeded to
calculate the $\beta$-function for a Yang-Mills theory. Today this calculation
is
regarded as quite simple and even  assigned as a homework problem  in
quantum field theory courses. At the time it was not so easy. This change in
attitude is the analogue, in theoretical physics, of the familiar phenomenon in
experimental physics whereby yesterday's great discovery becomes today's
background. It is always easier to do a calculation when you know what the
result is and you are sure that the methods make sense.

 One problem we had to face was that of gauge invariance. Unlike
QED, where the charge renormalization was trivially gauge invariant
(because the photon is neutral), the renormalization constants in QCD were all
gauge dependent. However the physics could not depend on the gauge.
Another issue was the choice of regularization. Dimensional regularization had
not really  been developed yet, and we had to convince ourselves that the
one-loop $\beta$-function was insensitive to the regularization used.
We  did the calculation in an arbitrary gauge. Since we knew
that the answer had to  be gauge invariant, we could use gauge invariance
as a check on our arithmetic.  This
was good since we both kept on making mistakes. In February the pace picked
up, and  we completed the calculation in a spurt of activity. At one point a
sign
error in one term convinced us that the theory was, as expected,
non-asymptotically free. As I sat down to put it all
together and to write up our results, I caught the error. At almost the same
time Politzer   finished his
calculation  and we compared, through Sidney, our results. The agreement was
satisfying.

A month or two after this Symansik passed through Princeton and
told us that 't Hooft had made a remark in a question session during a
meeting at Marseilles the previous fall to the effect that non-Abelian gauge
theories worked in the same way as an asymptotically free scalar theory he had
been playing with.\footnote*{This scalar theory was ruled out, as Coleman and I
argued \refmark{\grocol},
since one could prove it had no ground state and therefore was unstable.}
He did not publish and apparently did not realize the significance for scaling
and
for the strong interactions.

           Why are non-Abelian gauge theories asymptotically free? Today we
can understand this in a very physical fashion, although it was certainly not
so clear in 1973. It is instructive to interrupt the historical narrative and
explain, in
modern terms, why QCD is asymptotically free.

The  easiest way to understand this is by
considering the  magnetic screening properties of the vacuum
\refmark{\Niel}.In
a relativistic theory one can calculate the  dielectric constant,
$\epsilon$, in terms of the magnetic permeability,
 $\mu$, since  $\epsilon \mu=1$ (in units where $c$=velocity of light=1). In
classical physics all media are diamagnetic. This is because,  classically, all
magnets  arise from  electric currents and the response of a system to an
applied
magnetic field is to set up currents that act to decrease
the field (Lenz's law). Thus $\mu < 1$, a  situation that  corresponds to
electric screening or $\epsilon > 1$. However, in quantum mechanical systems
paramagnetism is possible. This is the case in non Abelian gauge
theories where the gluons are charged particles of spin one. They
behave as permanent color magnetic dipoles that align themselves
parallel to an applied external field increasing its magnitude
and producing  $\mu > 1$.  We can therefore regard the
anti-screening    of the Yang-Mills vacuum as paramagnetism!

                QCD is asymptotically free because the anti-screening of
the gluons  overcomes the screening due to the quarks. The arithmetic
works as follows. The contribution to $\epsilon$ (in some units)  from a
particle
of charge $q$ is $-  {q^2 \over 3}$, arising from ordinary dielectric
(or diamagnetic) screening. If the particle has spin $s$ (and thus a permanent
dipole
moment $\gamma s$), it  contributes $(\gamma s)^2$  to $\mu$. Thus a
spin one gluon (with $\gamma=2$, as in Yang-Mills theory) gives a contribution
to $\mu$ of
$$\delta \mu =(-1/3+2^2)q^2={11\over 3}q^2; $$
\noindent whereas a spin one-half quark contributes,
$$\delta \mu=-(-1/3+(2\times  {1\over 2})^2)q^2 =-{2 \over 3} q^2.
$$
\noindent(the extra minus arises because
quarks are fermions).
In any case, the upshot is that  as long as there are not too many
quarks the anti-screening of the gluons wins out over the screening
of the quarks.

The formula for the $\beta $-function of a non-Abelian gauge theory is
given by

$$ \beta (\alpha)\equiv \mu {d \over d \mu} \alpha(\mu)|_{\alpha_{\rm bare\,\,
fixed}} = {{\alpha}^2\over \pi} b_1 + ({{\alpha}^2\over \pi})^2 b_2 + \dots
; \qquad \alpha= {g^2\over 4 \pi}\,\,. \eqn\betaa  $$
Our result was that
$$
b_1 = -[ {11\over 6} C_A - {2\over 3}\sum_R n_RT_R ] \,\,. \eqn\form $$
Here $C_R$~is the eigenvalue of the quadratic Casimir operator
in the representation R of SU(N) (for the adjoint representation $C_A=N$,
for the fundamental $C_F={N^2-1\over N}$), $T_R$ is trace of the square
of the generators for the representation R  of SU(N) ($T_A$= N and $T_F= {1
\over 2}$), and $n_R$ is the number of fermions in the representation R.
In the case of a SU(3) gauge group such as QCD, $C_A=3$,
$T_F$=2, and thus $b_1= -[{11\over 2} - {n \over 3}]$. Thus   one can tolerate
as many as  16 triplets of quarks before losing asymptotic freedom.

\bigskip
\section{\bf Non-Abelian Gauge  Theories of the Strong Interactions}
\bigskip

For me the discovery of asymptotic freedom was totally unexpected.
Like an atheist who has just received a message from a burning bush, I became
an immediate true believer. Field theory wasn't wrong--instead
scaling must be explained by an asymptotically free gauge theory of the strong
interactions.
Our first paper contained, in addition to the report of the asymptotic freedom
of
Yang-Mills theory, the hypothesis that this could offer an explanation for
scaling,
a remark that there would be logarithmic violations of scaling  and most
important of all the suggestion that the strong interactions must be based on a
color gauge
theory.  The first paragraph reads \refmark{\growil}:

{\sl Non-Abelian gauge theories have received much attention
recently as a means of constructing unified and renormalizable theories of the
weak and
electromagnetic interactions. In this note we report on an investigation of the
ultraviolet asymptotic behavior of such theories. We have found that they
possess
the remarkable feature, perhaps unique among renormalizable theories, of
asymptotically approaching free-field theory. Such asymptotically free theories
will
exhibit, for matrix elements of currents between
on-mass-shell states, Bjorken scaling. We therefore
suggest that one should look to a non-Abelian gauge theory
of the strong interactions to provide the explanation for Bjorken scaling,
which has so far eluded field theoretic understanding."}

 We had a specific theory in mind. Since the deep-inelastic experiments
indicated
that the charged constituents of the nucleon were quarks, the gluons had to be
flavor
neutral. Thus the gluons could not couple to flavor.  We were very aware of the
growing
arguments for the color quantum
number. Not just the quark model spectroscopy that was the original motivation
of Greenberg \refmark{\green} and Han and Nambu
\refmark{\hannambu}, but the counting factor (of three) that went into the
evaluation of the $\pi^0\rightarrow 2 \gamma$ decay rate from the axial
anomaly\footnote* {This had been recently emphasized by William Bardeen,
Harold Fritzsch and
Gell-Mann\refmark{\bfgm}.}, and the factor of three that color provided in the
total
$e^+-e^-$ annihilation cross section. Thus the gluons could couple to color and
all
would be  well. Thus we proposed \refmark{\growil}:

{\sl   \lq\lq One particularly appealing model is based on three triplets of
fermions, with Gell-Mann's $SU(3)\times  SU(3)$ as a global symmetry and a
$SU(3)$
\lq color' gauge group to provide the strong interactions. That is, the
generators of the strong interaction gauge group commute with ordinary
$SU(3)\times
SU(3)$ currents and mix quarks with the same isospin and hypercharge but
different \lq color'. In such a model the vector mesons are (flavor) neutral,
and
the structure of the operator product expansion of electromagnetic or weak
currents is essentially that of the free quark model (up to calculable
logarithmic
corrections)."}

        The appearance of logarithmic corrections to scaling in asymptotically
free theories had already been discussed
by Callan and me, in our  work on the need for an asymptotically free
theory to obtain Bjorken scaling. We also analyzed deep inelastic
scattering in an asymptotically free theory and discovered \refmark{\grocall}

 {\sl   \lq\lq That in such asymptotically free theories naive scaling is
violated by calculable logarithmic terms." }

        Thus we were well aware what the form of the scaling deviations would
be
in such a theory. Wilczek and I had immediately  started to calculate the
logarithmic deviations from scaling. We had already evaluated the asymptotic
form of the flavor non-singlet structure functions, which were the
easiest to calculate, at the time our
{\it  Physical  Review Letter} was written, but did not have room to include
the
results. We immediately started to write a longer paper in which the structure
of the theory would be spelled out in more detail and the dynamical issues
would
be addressed, especially the issue of confinement. In our letter we were rather
noncommittal on this issue. We had tentatively concluded that Higgs mesons
would destroy asymptotic freedom, but had only begun to explore the dynamical
consequences of unbroken color symmetry. The only thing we were sure of was
that \refmark{\growil}

{\sl \lq\lq \dots perturbation theory is not trustworthy with respect to the
stability of
the symmetric theory nor to its particle content .\rq\rq }

Politizer's paper appeared with ours \refmark{\pol}. He pointed out the
asymptotic freedom of Yang-Mills theory and speculated on its implications for
the dynamical symmetry breaking of these theories.

  In our second paper, written a few months later, we outlined in
much greater detail the structure of asymptotically free gauge theories
of the strong interactions and the predictions for the scaling violations in
deep-inelastic scattering \refmark{\growill}.
Actually the paper  was
delayed for about two months because
we had problems with the singlet structure functions--due to the operator
mixing
of physical operators with ghost operators. This problem was similar to
the issue of gauge invariance that had plagued us before. Here the problem
was more severe. Physical operators, whose matrix elements were
measurable in deep-inelastic scattering experiments, mixed under
renormalization with ghost operators that could have no physical
meaning. Finally we deferred  the analysis of the singlet structure functions
to a
third paper \refmark{\growilll}, in which we resolved this issue. We showed
that, even though this mixing was real and unavoidable, the ghost operators
decoupled from physical measurements.

In the second paper we discussed in detail the
choice between  symmetry breaking and unbroken symmetry and noted that
\refmark{\growill}

{\sl \lq\lq  Another possibility is that the gauge symmetry is exact. At first,
 sight this would appear to be ridiculous since it would imply
the existence of massless, strongly coupled vector mesons. However, in
asymptotically free
theories these naive expectations might be wrong.
There may be little connection between the \lq free'
Lagrangian and the spectrum of states.....The infrared behavior of
Green's functions in this case is determined by the strong-coupling limit of
the
theory. It may be
very well that this infrared behavior is such so as to suppress all
but color singlet states, and that the colored gauge fields as well as the
quarks could be \lq seen' in the large-Euclidean momentum region but never
produced as real asymptotic states."}

 Steve Weinberg  reacted immediately to asymptotic freedom. He wrote a paper in
which he pointed out that in an asymptotically free gauge theory of the strong
interactions the non-conservation of parity and strangeness can be
calculated ignoring the strong interactions,
and thus is of order $\alpha$, as
observed. He also suggested that  a theory with unbroken color symmetry could
explain why we do not see quarks.

There is a slight difference between our respective conjectures.
Weinberg argued that perhaps the {\em infrared divergences}, caused by the
masslessness of the gluons in an unbroken color gauge theory, would make the
rate of  production of non-singlet states vanish. We argued that perhaps the
growth of the effective coupling at large distances, the {\em infrared
behavior} of
the coupling  caused by the flip side of asymptotic freedom\footnote*{Later
dubbed {\em infrared slavery} by Georgi and Glashow\refmark{\georglas}},
would confine the quarks and gluons in  color
singlet states.

In October 1973 Fritzsch, Gell-Mann and H. Leutwyler submitted a
paper in which they discussed the {\sl \lq\lq advantages of
color octet gluon picture" }\refmark{\fgml}.
Here they discussed the advantages of {\sl \lq\lq abstracting properties of
hadrons and their currents from a Yang-Mills gauge model based on colored
quarks and color
octet gluons." }They discussed various models and pointed out the
advantages of each. The first point was already discussed
at the NAL high energy physics conference in August 1972. There Gell-Mann
and Fritzsch had discussed their program of {\sl \lq\lq abstracting results
from the
quark-gluon model."} They discussed various models and asked, {\sl \lq\lq
Should
we regard the gluons as well as being color non-singlets."} They noted
that if one assumed that the gluons were color octets then {\sl \lq\lq an
annoying
asymmetry between quarks and gluons is removed."} In that talk no dynamical
theory was proposed and in most of the paper
they {\sl \lq\lq  shall treat the vector gluon, for convenience, as a color
singlet."}
\refmark{\nal}  In October 1973 Fritzsch, Gell-Mann and Leutwyler also noted
that
in the non-relativistic quark model with a Coulomb potential mediated by vector
gluons the potential is attractive in color singlet channels, which might
explain
why these are light. This point had been made previously by Harry Lipkin
\refmark{\lipkin}. They also noted the asymptotic freedom of such theories, but
did not regard this as an argument for scaling since {\sl \lq \lq we conjecture
that
there might be a modification at high energies that produces true scaling."}
Finally they noted that the axial U(1) anomaly in a non-Abelian gauge theory
might explain the notorious {\em U(1) problem}, although they could not explain
how, since the anomaly itself could be written as a total divergence.\footnote
\dagger{ It
required the discovery of instantons \refmark{\instant} to find the explanation
of the
{\em U(1) problem}. \refmark{\thooftinst , \thetaa 	}}

\bigskip
\chapter {\bf The Emergence and Acceptance of QCD}
\smallskip

Although it was clear to me that the strong interactions must be described by
non-Abelian gauge theories, there were many problems. The experimental
situation was far from clear, and the issue of confinement remained open.
However, within a small community of physicists the acceptance of the theory
was very rapid. New ideas in physics sometimes take years to percolate into
the collective consciousness. However in rare cases such as this there is
a change of perception analogous to a phase transition. Before asymptotic
freedom it seemed that we were still far from a dynamical theory of hadrons;
afterwards it seemed clear that QCD \footnote* {The name {\em QCD} first
appeared in a review by  Bill Marciano and Heinz Pagels \refmark{\pagels},
where it was attributed to Gell-Mann.  It was such an appropriate name that no
one could complain.}was such a theory.  Asymptotic freedom explained scaling
at  short distances and  offered  a mechanism for confinement at large
distance.
Suddenly it was clear that a non-Abelian gauge theory was consistent with
everything we knew about the strong interactions. It could encompass all the
successful strong interaction phenomenology of the past decade. Since the
gluons were flavor neutral, the global flavor symmetries of the strong
interactions,
SU(3)$\times$ SU(3), were immediate consequences of the theory, as long as
the masses of the quarks were small enough. \footnote\dagger{I refer  of course
to the
mass parameters of the quarks in the Lagrangian,  not the physical masses  that
are effectively infinite due to confinement.} Even more alluring was the fact
that
one could calculate. Since perturbation theory was trustworthy at short
distances
many problems could be tackled. Some theorists  were immediately convinced,
among them Guido Altarelli, Tom Appelquist, Callan, Coleman,
Mary K. Gaillard, R. Gatto, Georgi, Glashow, John Kogut, Ben Lee,
Luciano Maiani, Migdal, Polyakov,
Politzer, Lennie Susskind, S. Weinberg, Zee.

At large distances however perturbation theory was useless. In fact, even today
after nineteen years of study we still lack reliable, analytic tools for
treating this
region of QCD.  This remains one of the most important, and woefully neglected,
areas of theoretical particle physics. However, at the time the most important
thing was
to convince oneself that the idea of confinement was not inconsistent. One of
the
first steps in that direction was provided by lattice gauge theory.

 I first heard of Wilson's lattice gauge theory when I gave a lecture at
Cornell
in the late spring of 1973.  Wilson had started to think of this approach  soon
after asymptotic freedom was discovered. The lattice formulation of gauge
theory
(independently proposed by Polyakov) had the enormous advantage, as Wilson
pointed out in the fall of 1973, that the strong coupling limit was
particularly
simple and exhibited confinement\refmark{\wilcon}. Thus one had at least a
crude approximation
in which confinement was exact. It is a very crude approximation, since to
arrive
at the continuum theory from the lattice theory one must take the weak coupling
limit. However one could imagine that the property of confinement was not lost
as
one went continuously from  strong to weak lattice coupling, \ie~there was no
phase transition.  Moreover one could, as advocated by Wilson, study this
possibility numerically using Monte Carlo methods to construct the lattice
partition
function. However, the first quantitative results of this program did not
emerge till
the work of Creutz  \refmark{\Creuz} in 1981. The ambitious program of
calculating the hadronic  mass spectrum has still not attained its goal, and
still awaits
the next generation of computers.

        Personally I derived much solace in the coming year from two examples
of
soluble two dimensional field theories. One was the $(\bar \Psi \Psi)^2$ theory
that Neveu and I analyzed and solved for large N \refmark{\gronev}.
This provided a soluble example of an asymptotically free theory
that underwent dimensional transmutation, solving its infrared problems by
generating  a dynamical fermion mass through  spontaneous symmetry breaking.
This provided a model of an asymptotically free theory, with no built in mass
parameters. We could solve this model and check
that it was consistent and physical. The other soluble
model was two dimensional QCD, analyzed by t'Hooft in the large N
limit\refmark{\thoot}. Two dimensional gauge theories trivially confine color.
This
was realized quite early  and discussed for Abelian gauge theory--the Schwinger
model-- by Aharon Casher, Kogut and Susskind,  as a model for confinement  in
the fall
of 1973 \refmark{\casher}. However $QCD_2$ is a  much better example. It has a
spectrum of confined quarks which in many ways resembles the four
dimensional world. These examples  gave many of us total confidence in the
consistency of the concept of confinement. It clearly was possible to have a
theory whose basic fields do not correspond to asymptotic states, to
particles
that one can observe directly in the laboratory.

						Applications of the theory also began to appear. Two calculations of the
$\beta$-function to two loop order were performed \refmark{\cas,\Jones},
with the result that, in the notation of \form, $b_2=-[{17 \over 12} C_A^2 -
{1 \over 2}C_F T_F n - {5\over 6} C_A T_F n]$. Appelquist and Georgi and Zee
calculated
the corrections to the scaling of the $e^+-e^-$ annihilation cross section
\refmark{\appelt, \Zet}. Gaillard, and Lee \refmark{\MaryK}, and independently
Altarelli and Maiani\refmark{\Alt}, calculated the enhancement of the $\Delta
I={1\over 2}$ non-leptonic decay matrix elements.
The analysis of scaling violations for deep-inelastic scattering
continued \refmark{\scaletest}, and
the application of asymptotic freedom, what is now called
{\em perturbative QCD}, was extended to many new processes.

        The experimental situation developed slowly, and initially
looked rather bad. I remember in the spring of 1974  attending a meeting in
Trieste.  There I met Burt Richter who was gloating over the fact that $R=
{\sigma_{e^+ e^- \to \rm hadrons}\over \sigma_{e^+ e^- \to  \mu^+\mu^-}}$ was
increasing with energy, instead of approaching the expected constant value.
This was the most firm  of all the scaling predictions. $R$ must approach a
constant in any scaling theory. In most  theories however one cannot predict
the
value of the constant. However, in an asymptotically  free theory the constant
is
predicted to equal    the sum of the squares of the charges of the
constituents.
Therefore if there were only the three {\em observed} quarks, one would expect
that $R \to 3[({1\over 3})^2+ ({1\over 3})^2+  ({2\over 3})^2] =2 $.
However Richter reported that $R$ was  increasing, passing through $2$, with no
sign of flattening  out. Now many of us knew that charmed particles had to
exist.
Not only were they required, indeed invented, for the GIM mechanism to work,
but
as Claude Bouchiat, John Illiopoulos and Maini \refmark{\bim}and Roman Jackiw
and I
\refmark{\grojak} showed, if the charmed quark were  absent  the electro-weak
theory would be anomalous and  non-renormalizable.\refmark{\anom} Gaillard,
Lee and Jonathan Rosner had written an important and insightful paper on the
phenomenology of charm\refmark{\GaLeRos}.
Thus, many of us thought that since $R$ was  increasing
probably charm was being produced.

In 1974 the charmed mesons, much  narrower
than anyone imagined\footnote*{ except for Appelquist and
Politzer\refmark{\appelt}} were discovered, looking very much like
positronium--Coulomb bound states  of quarks. This clinched the matter for many
of the remaining skeptics. The rest were probably convinced once experiments at
higher energy began to see quark and gluon jets.

The precision tests of the theory, the logarithmic deviations from scaling,
took
quite a while to observe. I remember very well a remark made to me by a senior
colleague, in April of 1973 when I was very high, right after the discovery of
asymptotic freedom. He remarked that it was unfortunate that our  new
predictions regarding deep-inelastic scattering were  logarithmic effects,
since it
was unlikely that we would see them verified, even if true, in our lifetime.
This
was an exaggeration, but the tests did take a long time to appear. Confirmation
only started to trickle in in 1975 -78; and then at  a slow pace. By now the
predictions are indeed verified, in some cases to better than a  percent.

Nowadays, when you listen to experimentalists talk about
their results they point to their lego plots and say, \lq \lq
Here we see a quark, here a
gluon." Believing is seeing, seeing is believing. We now believe in the
physical
reality of quarks and gluons; we now believe in asymptotic simplicity of their
interactions at high energies so we can {\em see} quarks and gluons. The way in
which we see quarks and gluons, indirectly through the effects they have on our
measuring instruments, is  not much different from the way we see electrons.
Even the objection  that quarks and gluons can not be real particles, since
they
can never be isolated, has largely been dissipated. If we
were to heat the world to a temperature of a few hundred Mev, hadrons would
melt into a plasma of liberated quarks and gluons.

\bigskip

\chapter { \bf  Other Implications of Asymptotic Freedom.}
 \section{\bf Consistency of Quantum Field Theory}

Traditionally, fundamental theories of nature have had a tendency to break down
at short distances. This often signals the appearance of new physics that is
discovered once one has experimental instruments of high enough resolution
(energy) to explore the higher energy regime. Before asymptotic freedom it was
expected that any quantum field theory would fail at sufficiently high energy,
where the flaws of the renormalization procedure would appear. To deal with
this,
one would have to invoke some  kind of {\em fundamental length}.  In an
asymptotically free theory this is not necessarily the case--the decrease of
the
effective coupling for large energy means that no new physics need arise at
short
distances. There are no infinities at all, the bare coupling is finite--indeed
it
vanishes. The only divergences that arise are an illusion  that appears when
one
tries to compare, in perturbation theory,  the finite effective coupling at
finite
distances  with  the vanishing effective coupling at infinitely short
distances.

Thus the discovery of asymptotic freedom greatly reassured one of the
consistency of four-dimensional quantum field theory. One can trust
renormalization theory for an asymptotically free theory,  independent of the
fact
that perturbation theory is only an asymptotic expansion, since it gets better
and
better in the regime of short distances.  We are very close  to having a
rigorous mathematical proof of the existence of asymptotically free gauge
theories in four dimensions--at least when placed into a finite box to tame
the infrared dynamics that produces confinement. As far as we know, QCD by
itself is a totally consistent theory at all energies. Moreover, aside from the
quark
masses it has no arbitrary, adjustable parameters\footnote*{This is one of the
reasons it is so hard to solve.} Indeed, were it not for the electro-weak
interactions and gravity, we might be satisfied with QCD as it stands.

\bigskip
\section{\bf Unification}

        Almost immediately after the discovery of asymptotic freedom and the
proposal of the non-Abelian gauge theories of the strong  interactions, the
first
attempts were made to unify all the interactions.  This was natural, given that
 one
was using very similar theories  to describe all the known interactions.
Furthermore, the apparently insurmountable barrier to unification--namely the
large difference in the strength of the strong  interactions and the
electro-weak
interactions --was seen to be a low energy phenomenon. Since the strong
interactions decrease in strength with increasing energy these forces could
have
a common origin at very high energy. Indeed in the fall of 1974 Georgi and
Glashow proposed a unified theory, based on the gauge group $SU(5)$, which
remarkably contained the gauge groups of the standard model as well as the
quark and  lepton multiplets in an alluringly simple
fashion\refmark{\georglas}.
\footnote* {An earlier attempt to unify quarks and leptons was made by J. Pati
and Salam. \refmark{\Pati}} Georgi, Helen Quinn and Weinberg \refmark{\Quinn}
showed
that the couplings run in such a way as to merge somewhere around $10^{14}$
to $10^{16}$ Gev.

 This theory had the great advantage  of being tight enough to
make sufficiently precise  predictions  (proton decay and the Weinberg angle).
It
was a great stimulus for modern cosmology, since it implied that one could
extrapolate the standard model  to enormously high energies that
corresponded  to very early times in the history of the universe.
Although the $SU(5)$ theory has been invalidated by experiment, at least in its
simplest form, the basic idea that the next  fundamental  threshold of
unification is set by the scale where the strong and electro-weak couplings
become equal in strength remains at the heart of most attempts at unification.

\endpage
\refout

\end